\documentclass{article}
\usepackage{graphicx} % Required for inserting images
\usepackage{amsmath} % For cases environment

\title{A bend point analysis tool for the adequate truncation of city rank-size distributions}
\author{Rolf Bergs\thanks{PRAC, Bad Soden, Germany}}

\date{July 2026}

\begin{document}

\maketitle

\section{Introduction}
Zipf's law for cities is traditionally explained by a power law. There is, however, well substantiated evidence that the distribution follows Pareto only in the upper tail. Further down the curve it switches over to a lognormal regime, which visually appears to be hardly distinguishable from Pareto while its determination is completely different. A lognormal is not regularly varying compared to a Pareto distribution; therefore, a log-log Pareto regression over the entire distribution is essentially biased. This has motivated vital research on narrowing down the true break point or interval to best isolate the Pareto section. There have been several approaches to differentiate city rank-size distributions correctly along their Pareto and lognormal tail sections. Worth mentioning are log-likelihood ratio tests, UMPU, switching models, entropy based tests (Kullback-Leibler etc.), AIC/BIC parsimony or threshold point analysis based on the Hill estimator (inter alia: Malevergne et al 2011; Ioannides and Skouras 2013; Feng et al. 2020; Clauset et al 2009). The detection methods are largely complex, and, in comparison, not fully conclusive in their results. The method described here follows the estimation approach proposed by X. Gabaix and R. Ibragimov. It uses a logarithmic modification of the classical Zipf distribution to reduce bias in estimating the Pareto exponent of city sizes. Instead of ln(Rank), ln(Rank-1/2) is used (Gabaix and Ibragimov 2012). The modified Pareto form is transformed into the Zipf form. This way, the truncation method is less bias-affected and does not depend on more complex and - at times - less intuitive power estimations via maximum likelihood applied to avoid major bias as in traditional OLS power law regressions. The simple idea is that the largest positive difference between observed and estimated values indicates the greatest deviation of the empirical hierarchy from the ideal Zipf rule (\(\alpha \approx{1}\)) and thus constitutes a break point, the so called "hockey stick", typically found in urban scaling law studies (Chen et al. 2025). The important assumption is the typical quasi-concave curvature of the total distribution, where ideally the straight line connecting observation 1 with observation n (baseline regression) lies entirely below the distribution.
\\
\\
The method is exclusively determined from theory and a simulation exercise with a combined random generated hybrid distribution (upper tail: Pareto, lower tail: lognormal). Like the other truncation methods mentioned above, this one also ignores the fact that in practice two different distributions can also overlap continuously, so that the hybrid distribution does not break at a certain threshold value or interval, but large sections are defined by two or more distributions simultaneously. By definition, the double Pareto-lognormal distribution (DPLN), probably reflecting the real fractal world in urban systems, has no hard threshold; it is infinitely differentiable everywhere while its local log-log slope (the elasticity) changes continuously along the size classes (Reed and Jorgensen 2006). When considering functional urban clusters (metropolitan regions, density-based agglomerations) instead of simple administrative population data (cities proper), these sharp distinctions become blurred. The mix of centrality levels (i.e. cities of same size but different centrality level) appears as a continuous process of spatial structuring. The more fluid the geographic definition, the more likely the real system is to follow a continuous overlap (DPLN) rather than an abrupt spline model. This complex behavior cannot be precisely detected either by the methods mentioned above or by the fairly straightforward approach introduced here. The actual purpose of this exercise is to share a simple Stata thresholding tool (zipfbend.ado) for hybrid Pareto-lognormal distributions, notably to support studies addressing Zipf's law by adding a truncation method inspired by a simple but intuitive geometric explanation. 
\\
\\
\section{A simple model}
If modeling Zipf's law, i.e. Size=f(Rank) by exploring an upper randomly generated ranked Pareto tail with smoothly appended lower lognormally distributed random values (also ranked), then the largest positive deviation of observations from the log-log fitted regression line is essentially the transition point between the two tails (confirmed by numerous simulations). A hybrid (or composite) distribution with a lower lognormal body and an upper Pareto tail is thus formally defined as a spliced probability model. It uses the lognormal distribution to model the majority of smaller observations, and transitions to a heavy-tailed Pareto distribution at a detected threshold. 
\\
\\
When log-transforming a lognormal variable, it yields a normal distribution. In the log-transformed probability density function (PDF), the quadratic term dictates the parabolic curvature of the distribution when plotted on a log-log scale. It is responsible for tails thinning out rapidly compared to power-law distributions. The log-transformed density function of a lognormal variable is mathematically expressed as \(\ln(f(x)) = -\frac{(\ln x - \mu)^2}{2\sigma^2} - \ln(x\sigma\sqrt{2\pi})\). The quadratic component \(-\frac{(\ln x)^{2}}{2\sigma ^{2}}\) drives the shape of the distribution. Its role in the curvature is as follows: As the distance between \(\ln x\) and the mean \(\mu \) increases, the quadratic term grows. This causes the log-log plot of the PDF to curve downward into the tails, distinguishing it from power-law distributions which remain linear. There is also a variance impact. The denominator of the quadratic term depends on \(\sigma ^{2}\) (the variance of the underlying normal distribution). A larger \(\sigma ^{2}\) reduces the impact of the quadratic term, making the distribution appear linear on a log-log scale for a broader range of values (Mitzenmacher and Tworetzky 2003).
\\
\\
To explain the above mentioned simulation findings mathematically and to substantiate the computing approach, one must explore the functional form of the residual between the true merged distribution and the linear regression model. The largest positive deviation occurs at the transition point because it marks the maximum structural divergence between the quadratic (lognormal) regime and the linear (Pareto) regime on a log-log scale. Let \(x\) be the log-rank (\(x = \ln (R-0.5)\)) and \(y\) be the log-size (\(y = \ln S\)):
\\
\\
For the largest entities (low ranks, small \(x\)), the system follows a true power law (Zipf's law). In log-log space, this is strictly linear, i.e. \(y_{P}(x)=C_{1}-\alpha x\), where \(\alpha \) is the Zipf coefficient (reverse Pareto).
\\
\\
For smaller entities (high ranks, large \(x\)), the distribution switches to lognormal. The survival function of a lognormal variable in the log-log space contains a dominant negative quadratic term \(y_{logn}(x)=C_{2}-\beta _{1}x-\beta _{2}x^{2}\), where \(\beta_2 > 0\), giving it a strictly concave curvature. By smoothly appending the lognormal tail to a transition point \(x_0 = \ln R_0\), the true data-generating process \(y(x)\) is continuous and differentiable:
\\
\\
\(y(x)=\begin{cases}C_{1}-\alpha x&\text{for\ }x\le x_{0}\\ C_{2}-\beta _{1}x-\beta _{2}x^{2}&\text{for\ }x>x_{0}\end{cases}\). In this way, we arrive at a merged model.
\\
\\
When we fit a standard Ordinary Least Squares (OLS) linear regression model \(\hat{y}(x) = a + bx\) across the entire dataset, the regression line acts as a chord that intersects the overall curve. Because the lognormal body dominates the sample size (the lower body contains the vast majority of observations in empirical systems), the OLS fit is heavily weighted by the concave lognormal section. This pulls the regression line \(\hat{y}(x)\) downward in the lower tail to balance the quadratic drop-off. As a result, the fitted slope \(b\) is steeper (more negative) than the true Pareto tail slope \(-\alpha \), namely: \(|b|>\alpha \implies b<-\alpha \).
\\
\\
The deviation (residual) at any point \(x\) is defined as: \(D(x)=y(x)-\hat{y}(x)\). To find the maximum positive deviation, we analyze the behavior of the derivative \(D'(x)\) on both sides of the transition point \(x_{0}\).
\\
\\
Left of the Transition (\(x \leq x_0\)): In the Pareto tail, the deviation is the difference between two linear functions:
\\
\\
\(D(x)=(C_{1}-\alpha x)-(a+bx)=(C_{1}-a)+(b-\alpha )x\). 
\\
\\
Taking the first derivative with respect to \(x\): \(D^{\prime }(x)=b-\alpha \). Since \(b < -\alpha\), the term \((b - \alpha)\) is strictly positive (\(D'(x) > 0\)). This means that throughout the entire Pareto tail, the positive deviation increases monotonically as one moves toward \(x_{0}\).
\\
\\
Right of the Transition (\(x > x_0\)): Immediately after crossing into the lognormal regime, the quadratic term takes over:
\\
\\
\(D(x)=(C_{2}-\beta _{1}x-\beta _{2}x^{2})-(a+bx)\)
\\
\(D^{\prime }(x)=-\beta _{1}-2\beta _{2}x-b\). 
\\
\\
Due to the smooth-appending constraint (first-order differentiability at \(x_{0}\)), the derivatives must match at the boundary: \(y'_{logn}(x_0) = y'_{P}(x_0) = -\alpha\). Therefore, right at the transition \(D^{\prime }(x_{0}^{+})=-\alpha -b>0\). However, as \(x\) increases beyond \(x_{0}\), the \(-2\beta_2 x\) term grows larger and more negative. The derivative quickly changes sign from positive to negative (\(D'(x) < 0\)), driving the deviation downward into negative territory as the lognormal curve plunges below the regression line.
\\
\\
Because \(D'(x) > 0\) for all \(x < x_0\) (strictly increasing approach) and becomes negative shortly after \(x_{0}\) due to the quadratic deceleration of the lognormal component, the global maximum of the residual function must locate exactly where the regime shift occurs, namely: \(\max _{x}D(x)\approx D(x_{0})\). The simulations thus align with this calculus: the transition point is mathematically hardwired to be the apex of the positive residual because it represents the precise threshold where the data stops being perfectly linear and begins to decelerate quadratically.
\\
\\
Summary of the Proof: In theory, the transition point \(x_{0}\) yields the largest positive deviation from the fitted regression line. A positive residual also means that, relative to its size, a city ranks higher (i.e., is larger) than the ideal Zipf distribution would predict for that position in the system. The slope evolves because the OLS line is pulled down by the concave lognormal tail, making its slope steeper than for the true Pareto tail. This causes the residual to steadily grow throughout the Pareto region, hitting its peak exactly at the boundary \(x_{0}\) before the quadratic decay of the lognormal distribution forces the residual to collapse (see figure 1). This, however, only applies to a perfectly shaped spline model, where all observations are covered in the estimation.
\\
\\
An important limitation is the assumption of a strictly linear distribution of the logged Pareto tail and strictly monotonically decreasing logarithmic values in the lower (log-normal) tail. Even a small local variation in the distribution can indicate an incorrect transition point. Because the estimate depends on the specific segment of the distribution under consideration, the identification of the truncation point is thus not always unambiguous, since in real world scenarios multiple maxima of the deviations may occur, just depending on the truncation and the respective position of the regression line. This calls for an extended and more robust method to detect the transition point (see figure 1).

\begin{figure}
    \centering
    \includegraphics[width=1\linewidth]{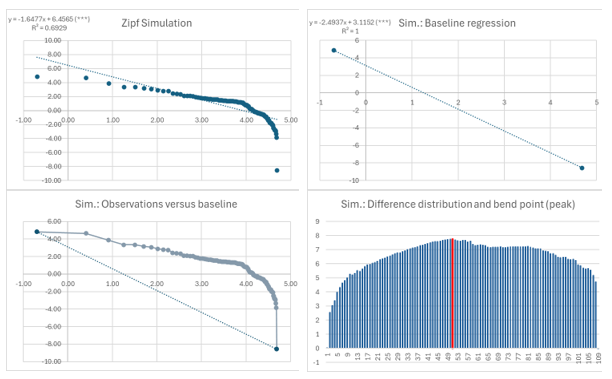}
    \caption{Simulation of a hybrid Pareto-lognormal distribution}
    \label{fig:placeholder}
\end{figure}

\section{Definition of variables and the regression approach}
The method builds on two simple log-linear reversed Pareto models with:

\[ln(y) = a - b \cdot ln(x)+e\]
\\
to estimate the varying power exponents depending on the truncation point, and additionally a stable "chord" (secant) as a baseline linking the first (1) and the last (n) observation:
\\
\[\ln(y_i) = a - b \cdot \ln(x_i) + e_i \quad \text{for} \; i \in \{1, n\} \] to find the maximum difference between the stable baseline and the logged observations, regardless of which segments of the distribution are being examined. In both models y means size (e.g. population) and x means rank-1/2. 
\\
\\
Then we solve the baseline equation for \(x\) to calculate the theoretically expected rank for each city size \(y\): \(x_{estimated} = \frac{y - a}{-b}\)
For each rank we calculate the difference: \(\Delta x = x_{\text{observed}} - x_{\text{estimated}}\). The largest positive difference from the baseline is the maximum value in this series: \(\max(\Delta x)\). This maximum positive value may thus indicate a bend point in the function, where to locate a transition point between Pareto and lognormal irrespective of a changing overall coefficient effected by alternative truncations. The simulation exercise with a pre-defined threshold clearly shows the bend point at that rank. The specific advantage of the baseline regression is its stable position that does not vary with the truncation point. As said earlier, in the real world of cities, not all local bend points are true transition points, and this issue still cannot be fully solved by the use of the stable secant. Several apparent transition points may originate from the variance of neighboring city sizes along the ranks (discrete jumps instead of continuous behavior, partly effected by overlapping distribution types); further to that, the primate city may also exert disturbing influence if it is substantially larger than double the size of the second ranked city (with the possible implication of a convex top section). The often observed overwhelming weight of the primate city suggests either to winsorize the distribution or even to delete the largest city from the sample. Then, to detect the true (or most adequate) transition point, the distribution of the regarded section needs to be tested and interpolated by truncating sections until a true Pareto distribution remains. A useful test for that is Kolmogorov-Smirnov. The usual approach would be thus to first run the procedure for all observations. If then H0 (data follows Pareto) is rejected, the range between rank one and the detected bend point rank can be viewed. This will indicate a new bend point located more in the upper tail. But if then H0 were accepted (pure Pareto is plausible), the former bend point could be taken as an adequate truncation rank.
\\
\\
\section{References}
\noindent Chen C, Zhang X, Webster C (2025) Truncation in the scaling of urban pollution control. NPJ Urban Sustainability 5:9
\\
\\Clauset A, Shalizi C, Newman MEJ (2009) Power-Law Distributions in Empirical Data. SIAM Review 51:661-703
\\
\\Feng M, Deng LJ, Chen F, Perc M, Kurths J (2020) The accumulative law and its probability model:an extension of the Pareto distribution and the log-normal distribution.Proceedings of the Royal Society A 476:2020001
\\
\\Gabaix X, Ibragimov R (2012) Rank - 1/2: A Simple Way to Improve the OLS Estimation of Tail Exponents, Journal of Business and Economic Statistics, 29:1, 24-39
\\
\\Ioannides Y, Skouras S (2013) US city size distribution: Robustly Pareto, but only in the tail. Journal of Urban Economics 73: 18-29
\\
\\Malevergne Y, Pisarenko V, Sornette D (2011) Testing the Pareto against the lognormal distributions with the uniformly most powerful unbiased test applied to the distribution of cities. Physical Review E 83: 036111
\\
\\Mitzenmacher M, Tworetzky B (2003) New methods and models for file size distributions. Mimeo: Harvard University: 
\\https://www.eecs.harvard.edu/~michaelm/postscripts/aller2003.pdf
\\
\\Reed WJ, Jorgensen M (2006) The Double Pareto-Lognormal Distribution—A New
Parametric Model for Size Distributions. Communications in Statistics - Theory and Methods, 33:8, 1733-1753
\\
\\
\\
\\
\\
\\
\section{Annex: The Stata ado file: "zipfbend"}
zipfbend \textit{varname} [\textit{if}] \textit{in} [\textit{, options}]
\\
\\
The range of observations needs to be specified. This makes it possible to compare the behavior of different sections along the distribution. Options include [\textit{if}] and the generation of a new variable containing the residuals, e.g. [\textit{, generate (k)}].
\\
\\
The results displayed comprise the estimated Zipf coefficient, the implied Pareto exponent, the maximum positive difference, its rank and size, and the Kolmogorov-Smirnov Test (Goodness-of-Fit for Power Law). If no option is selected, the command "return list" displays the various scalars saved.
\\
\\
Code download: https://github.com/bergsr/Stata

\end{document}